\documentclass[12pt]{article}
\usepackage[utf8]{inputenc}
\usepackage{graphicx}
\usepackage[utf8]{inputenc}
\usepackage{amsmath}
\usepackage{amsfonts}
\usepackage{amssymb}
\usepackage{graphicx}
\usepackage{setspace}
\usepackage{authblk}
\usepackage{caption}
\usepackage{float}
\usepackage{cite}
\usepackage[left=2cm,right=2cm,top=2cm,bottom=2cm]{geometry}
\usepackage{color}
\usepackage{color,soul}
\usepackage{caption}
\usepackage{subcaption}
\title{\textbf{A Non-Linear Type Equation of State and Cosmic Fluid Dynamics}}

\author[1]{Shouvik Sadhukhan}
\author[2]{Alokananda Kar}
\author[3]{Surajit Chattopadhyay}

\affil[1]{\small{Department of Physics; Indian Institute of Technology, Kharagpur 721302, West Bengal, India\par Email: shouvikphysics1996@gmail.com}}
\affil[2]{\small{Department of Physics; University of Calcutta; 92 APC Road, Kolkata 700009, West Bengal, India \par Email: {alokanandakar@gmail.com}}}
\affil[3]{\small{Department of Mathematics; Amity University; Newtown; Kolkata; West Bengal; India\par corresponding author Email: surajitchatto@outlook.com}}

\date{}
\begin{document}
\maketitle

\begin{abstract}
In this chapter we have introduced a special type of non-linear equation of state to discuss the cosmological evolution mechanism. The new equation of state is a four parameters model which can be represented as $p=A\rho+B\rho^2-\frac{C}{\rho^{\alpha}}$ where $B=A\beta-\gamma$. The evolution of universe have been interpreted  by fluid dynamics. The reconstruction of Chaplygin gas and Van-Der-Waals (VDW) fluid equation of states have been done from the parametric analysis of this new non-linear model. Different cosmological phases like Quintom, Quintessence and warm universe have been discussed here. Finally, we have provided a comparative studies of this model with other non-linear fluid solutions.\\

\end{abstract}

\section{Introduction}
Recent advancement of experimental cosmology provide the proof behind the accelerated expansion of universe. Type Ia Supernovae and the CMBR radiation analysis re-established the idea of cosmic inflation and late time acceleration scheme (Jamil.et.al (2009), Biswas.et.al (2013)). Those experimental results brought several questions of well known FRW models. The Horizon problem, Magnetic monopole problem and Flatness problems are few examples of them (Linde.et.al (1982)). For the resolution of these problems the cosmic inflation and the late time acceleration models have been brought in physics (Liddle.et.al (1999), Lucchin.et.al (1985)). The gravity provides attraction in local scale. Hence, to bring the repulsive nature, the idea of dark energy has got much attention. Dark energy is considered to act as exotic cosmic system that exerts repulsion against the  attractive gravity (Copeland.et.al (2006), Steinhardt.et.al (2003), Hughes.et.al (2019), Kar.et.al (2021, 2022)).\par
There are several dark energy models available in literature (Hoyel.et.al (1964), Zeldovich.et.al (1968), Capozziello.et.al (2006), Elizalde.et.al (2005)). The most acceptable models are Chaplygin gas type fluid dark energy models, scalar field type dark energy models and holographic dark energy models. There are several varieties of fluid type dark energy models viz. the chaplygin gas model (Kotambkar.et.al (2021)), modified chaplygin gas (Mamon.et.al (2022), Jianbo.et.al (2009)), cosmic chaplygin gas, modified cosmic chaplygin gas (Benaoum.et.al (2012), Khadekar.et.al (2019)), polytropic type fluid, modified polytropic or Van-Der-Waals fluid (Faraoni.et.al (2003), Capozziello.et.al (2002, 2003), Vardiashvili.et.al (2017), Ivanov.et.al (2019), Obukhov.et.al (2018), Jantsch.et.al (2016)) and Inhomogeneous fluids (Jamil.et.al (2008), Chakraborty.et.al (2008)). These fluid based dark energy models can show the self interacting dark matters which can substitute the dark energies through Cardassian formalism (Gondolo.et.al (2003), Frees.et.al (2002)). The inhomogeneous models can establish the theory of dissipation of self interaction of dark matters and can give a pause in the inflationary universe to get late time accelerated expansion phase (Kar.et.al (2021, 2022)). \par
The non-linear cosmological models are based on the non-linear pressure density relations or non linear equation of states. The non-linearity of equation of states are generally found in the introduction of self interaction (Nojiri.et.al (2004, 2006), Chakraborty.et.al (2019)). The evolutionary mechanism which follows those self interactions, are generally called Cardassian expansions. Those self interactions in Cardassian expansion scheme provides the proof of existence of fifth force or fifth dimensions (Nojiri.et.al (2004, 2006), Chakraborty.et.al (2019)). The decay of those fifth force can bring the graceful exit of inflation and start reheating (Albrecht.et.al (1982)) for late time accelerations. The non-linearities in equation of state are basically a combinations of linear dark matters and some non-linear fifth force based potentials which are active in Hubble scale but negligible in local scale. This force causes accelerated expansion against the gravitational collapsing (Kar.et.al (2021, 2022)).\par
In this present work we have used a new type of four parameters equation of states. This equation of state can be a generalization of both Chaplygin gas (Jianbo.et.al (2009)) and VDW fluids. This fluid model can be derived using binary type self interaction between linear dark energy, polytropic fluids and chaplygin gas (Elizalde.et.al (2018), Brevik.et.al (2018), Wu.et.al (2008)). We have only discussed the fluid dynamical properties and its effect on universe expansion into this paper. We have derived evolution scheme of the energy density and pressure of this model under Einstein gravity with non changing gravitational and cosmological constants. Finally, we have re-established the chaplygin gas (Benaoum.et.al (2012), Pourhassa.et.al (2014), Kotambka.et.al (2021)) and VDW fluid from this model along with the discussion of some cosmological phases.\par
The paper is oriented as follows. In section 2 we have discussed the mathematical basis of our work. In sections 3 and 4 we have given the reconstructions of different fluid models and different cosmological phases respectively. Finally, in sections 5 and 6 we have provided the results analysis and concluding remarks of our work.

\section{Mathematics on the New Non-linear fluid}
In this present work we want to study a new type of non-linear fluid model which can replace linear dark matter to discuss the expansion scheme of universe. We have used the action of constant gravitational constant with absence of cosmological constant type dark energies. We have taken $8\pi G=c=1$. The necessary action for our model can be written as follows.
\begin{equation}
    S=\int{d^4x(\sqrt{-g}R+L_f)}
\end{equation}
Now this action has been used on the isotropic and homogeneous geometry i.e., the FRW geometry. The line element has been taken as follows.
\begin{equation}
    ds^2=dt^2-a^2(t)(dr^2+r^2 d\Omega^2)
\end{equation}
The line element represents a flat universe and $d\Omega^2=d\theta^2+\sin^2{\theta}d\phi^2$. Now using least action principle on the action given above, we can find the following dynamical equation of cosmic mechanics i.e. the Einstein field equation.
\begin{equation}
    R_{\mu\nu}-\frac{1}{2}g_{\mu\nu}R=\kappa T_{\mu\nu}
\end{equation}
Here, $\kappa=1$ in SI unit. The term $T_{\mu\nu}$ comes from the cosmic matter Lagrangian $L_f$. The relation can be explained as follows.
\begin{equation}
    T_{\mu\nu}=-\frac{2}{\sqrt{-g}}\frac{\partial (\sqrt{-g}L_f)}{\partial g^{\mu\nu}}
\end{equation}
The tensor representation of the energy momentum tensor can be written as follows using the knowledge of isotropic energy distribution.
\begin{equation}
    T_{\mu\nu}=(\rho_f+p_f)u_{\mu}u_{\nu}-p_fg_{\mu\nu}
\end{equation}
Now, using the isotropic and homogeneous line element into the Einstein field tensor, we can find a couple of dynamical equations that can discuss the evolution of isotropic and homogeneous universe. The couple equations are called FRW equations which are as follows.
\begin{equation}
    3H^2=\rho_f
\end{equation}
And,
\begin{equation}
    3H^2+2\dot{H}=-p_f
\end{equation}
For any cosmological models, the fluid energy must be the conserved, i.e energy momentum tensor must be divergenceless. Similarly, the geometry must follow the divergenceless condition also. Hence we must have to satisfy $\nabla_{\mu}G_{\mu\nu}=\nabla_{\mu}T_{\mu\nu}$. Thus, we found the following energy momentum conservation equation.
\begin{equation}
    \dot{\rho}_f+3H(\rho_f+p_f)=0
\end{equation}
Now the solution of cosmic matter energy density and pressure should follow the energy conservation equation. The energy momentum conservation equation can provide the scale factor dependent variations of energy density and pressure which again can provide the time variation of scale factor by substitution of energy density and pressure from FRW equations. During the derivation of scale factor variations of energy density and pressure we must discuss the equation of state or pressure density relations of our model. This is given as follows:
\begin{equation}
    p_f=A\rho_f+B\rho_f^2-\frac{C}{\rho_f^{\alpha}}
\end{equation}
The non-linearity of the equation of state can provide dimensionally non unit parameters. Here, the dimensions of $A$, $B$ and $C$ are $[1]$, $[\rho]^{-1}$ and $[\rho]^{\alpha+1}$ respectively. The parameter $B$ can be dissociated as $B=A\beta-\gamma$ where the dimensions of $\beta$ and $\gamma$ are same as $B$. Now using the equation of state into the energy momentum conservation equation, we can find the following integral.
\begin{equation}
    \int{\frac{\rho_f^{\alpha}d\rho_f}{B\rho_f^{\alpha+2}+(A+1)\rho_f^{\alpha+1}-C}}+C_1=-3\ln{a}
\end{equation}
We have to find the solution of this equation which depends on four fluid parameters $A$, $\beta$, $\gamma$, $C$ and one power parameter $\alpha$. Hence, we have applied the fluid dissociation scheme to solve the energy density and pressure of our model. In the fluid dissociation method we have dissociated the fluid equation of state into two parts. One is $p_{f_1}=A\rho_{f_1}-\frac{C}{\rho_{f_1}^{\alpha}}$ and another one is $p_{f_2}=(A\beta-\gamma)\rho_{f_2}^2$. Here, the first part of equation of state act as cosmic fluid on p-brane whereas the second part of EOS act as the D-brane or fifth force based self interaction between the cosmic fluid particles. This dissociation is viable in physics as this second component works as fifth force to modify the effective pressure but don't work as second fluid system. Hence, the energy momentum tensor can also be dissociated into two equations without having any interaction scenario. The self interaction can not be shown with usual interaction scenario available in cosmology literatures. Hence, the effective energy density and pressures can be written as follows.
\begin{equation}
    \rho_f=\rho_{f_1}+\rho_{f_2}
\end{equation}
And,
\begin{equation}
    p_f=p_{f_1}+p_{f_2}=A\rho_{f_1}+(A\beta-\gamma)\rho_{f_2}^2-\frac{C}{\rho_{f_1}^{\alpha}}
\end{equation}
The energy momentum conservation equations for the dissociated system can be written as follows. 
\begin{equation}
    \dot{\rho_{f_1}}+3H(\rho_{f_1}+p_{f_1})=0
\end{equation}
And,
\begin{equation}
    \dot{\rho_{f_2}}+3H(\rho_{f_2}+p_{f_2})=0
\end{equation}
Now we solve those energy densities and pressures as a function of scale factor with usual ways. Hence, The solutions are as follows.
\begin{equation}
    \rho_{f_1}=(\frac{C}{1+A}+\frac{C_1}{a^{3(1+A)(1+\alpha)}})^{\frac{1}{1+\alpha}}
\end{equation}
\begin{equation}
    \rho_{f_2}=\frac{1}{C_1 a^3-(A\beta-\gamma)}
\end{equation}
\begin{equation}
    p_{f_1}==A((\frac{C}{1+A}+\frac{C_1}{a^{3(1+A)(1+\alpha)}})^{\frac{1}{1+\alpha}})-C(\frac{C}{1+A}+\frac{C_1}{a^{3(1+A)(1+\alpha)}})^{-\frac{\alpha}{1+\alpha}}
\end{equation}
And,
\begin{equation}
    p_{f_2}=(A\beta-\gamma)(\frac{1}{C_1 a^3-(A\beta-\gamma)})^2
\end{equation}
Hence, the effective energy density and pressures can be written using the above relations. The effective density is as follows.
\begin{equation}
    \rho_f=(\frac{C}{1+A}+\frac{C_1}{a^{3(1+A)(1+\alpha)}})^{\frac{1}{1+\alpha}}+\frac{1}{C_1 a^2-(A\beta-\gamma)}
\end{equation}
And, the effective pressure can be written as follows.
\begin{equation}
    p_f=A((\frac{C}{1+A}+\frac{C_1}{a^{3(1+A)(1+\alpha)}})^{\frac{1}{1+\alpha}})-C(\frac{C}{1+A}+\frac{C_1}{a^{3(1+A)(1+\alpha)}})^{-\frac{1}{1+\alpha}}+(A\beta-\gamma)(\frac{1}{C_1 a^3-(A\beta-\gamma)})^2
\end{equation}
Hence, the effective equation of state parameter can be written as follows.
\begin{equation}
\begin{split}
    \omega_f=\frac{p_f}{\rho_f} =\frac{A((\frac{C}{1+A}+\frac{C_1}{a^{3(1+A)(1+\alpha)}})^{\frac{1}{1+\alpha}})-C(\frac{C}{1+A}+\frac{C_1}{a^{3(1+A)(1+\alpha)}})^{-\frac{1}{1+\alpha}}+(A\beta-\gamma)(\frac{1}{C_1 a^2-(A\beta-\gamma)})^2}{(\frac{C}{1+A}+\frac{C_1}{a^{3(1+A)(1+\alpha)}})^{\frac{1}{1+\alpha}}+\frac{1}{C_1 a^3-(A\beta-\gamma)}}
    \end{split}
\end{equation}

Now using the above equations and controlling the parameters $A$, $\beta$, $\gamma$, $C$ and $\alpha$ we can reconstruct the chaplygin model solutions as well as VDW fluid solutions for cosmic evolution. Different phases and their smooth transformations through entire cosmic era can be discussed using the values of equation of state parameters.

\section{Generation of different fluids}
This new model discussed in the above section can be used to reconstruct different cosmological non-linear models with some parameter analysis. We can discuss the Chaplygin gas, Modified Chaplygin gas, Van-Der-Waals fluid and Polytropic fluid. The models can be reconstructed as follows.

\subsection{Chaplygin Gas}
For reconstructing the equation of state of Chaplygin gas, we have to choose the parameters as $A\rightarrow 0$, $(A\beta-\gamma)\rightarrow 0$ and $\alpha\rightarrow 1$. Hence, the equation of state should be as $p_f=-\frac{C}{\rho_f}$, pressure, density and EOS parameter $\omega_f$ can be written as follows.
\begin{equation}
    \rho_f=(\frac{C_1}{a^{6(1+A)}})^{\frac{1}{2}}
\end{equation}
\begin{equation}
    p_f=-C(\frac{C_1}{a^{6(1+A)}})^{-\frac{1}{2}}
\end{equation}
and,
\begin{equation}
    \omega_f=\frac{-C(\frac{C_1}{a^{6(1+A)}})^{-\frac{1}{2}}}{(\frac{C_1}{a^{6(1+A)}})^{\frac{1}{2}}}
\end{equation}
Hence, these results can produce pure Chaplygin gas and different cosmic phases from it.

\subsection{Modified Chaplygin Gas}
Now we want to reconstruct the equation of state of Modified Chaplygin gas from our model. For this purpose we must have to consider $(A\beta-\gamma)\rightarrow 0$. Hence, the equation of state can be taken as $p_f=A\rho_f-\frac{C}{\rho_f^{\alpha}}$. The pressure, density and EOS parameter $\omega_f$ can be written as follows.
\begin{equation}
    \rho_f=(\frac{C}{1+A}+\frac{C_1}{a^{3(1+A)(1+\alpha)}})^{\frac{1}{1+\alpha}}
\end{equation}
\begin{equation}
    p_f=A((\frac{C}{1+A}+\frac{C_1}{a^{3(1+A)(1+\alpha)}})^{\frac{1}{1+\alpha}})-C(\frac{C}{1+A}+\frac{C_1}{a^{3(1+A)(1+\alpha)}})^{-\frac{\alpha}{1+\alpha}}
\end{equation}
and,
\begin{equation}
    \omega_f=\frac{A((\frac{C}{1+A}+\frac{C_1}{a^{3(1+A)(1+\alpha)}})^{\frac{1}{1+\alpha}})-C(\frac{C}{1+A}+\frac{C_1}{a^{3(1+A)(1+\alpha)}})^{-\frac{\alpha}{1+\alpha}}}{(\frac{C}{1+A}+\frac{C_1}{a^{3(1+A)(1+\alpha)}})^{\frac{1}{1+\alpha}}}
\end{equation}

\subsection{Van-Der-Waals Fluid with $\left |\beta\rho_f \right |< 1$}
The reconstruction of Van-Der-Waals fluid can be done using the assumptions of parameters as $C\rightarrow 0$. Hence, the equation of state should become $p_f=A\rho_f+(A\beta-\gamma)\rho_f^2$ which is a special form of VDW model with condition of $\left |\beta\rho_f\right |<1$ (details is given in appendix). The mathematical solutions of energy density and pressure as well as equation of state parameter can be written as follows.
\begin{equation}
    \rho_f=\frac{1+A}{(\gamma-A\beta)+\rho_{f0}(1+A)a^{3(1+A)}}
\end{equation}
\begin{equation}
    p_f=A(\frac{1+A}{(\gamma-A\beta)+\rho_{m0}(1+A)a^{3(1+A)}})+(A\beta-\gamma)(\frac{1+A}{(\gamma-A\beta)+\rho_{m0}(1+A)a^{3(1+A)}})^2
\end{equation}
and,
\begin{equation}
    \omega_f=\frac{A(\frac{1+A}{(\gamma-A\beta)+\rho_{f0}(1+A)a^{3(1+A)}})+(A\beta-\gamma)(\frac{1+A}{(\gamma-A\beta)+\rho_{m0}(1+A)a^{3(1+A)}})^2}{\frac{1+A}{(\gamma-A\beta)+\rho_{f0}(1+A)a^{3(1+A)}}}
\end{equation}
This model can discuss the cosmic phases and their evolutions with three parameters. For this model we found some constraints on the parameters and scalar factors which can be explained as follows. We can assume that the scale factor can run only within the limit (see Appendix)
\begin{equation}
    (\frac{(1+A)\beta-(\gamma-A\beta)}{\rho_{f0}(1+A)})^{\frac{1}{3(1+A)}} < a < (\frac{(1+A)\beta+(\gamma-A\beta)}{\rho_{f0}(1+A)})^{\frac{1}{3(1+A)}}
\end{equation}
Similarly we can provide the limits of pressure of the activity of that equation of state as follows.
\begin{equation}
    -\frac{A}{\beta}+\frac{(A\beta-\gamma)}{\beta^2} < p_f < \frac{A}{\beta}+\frac{(A\beta-\gamma)}{\beta^2}
\end{equation}
Hence, pressure can be negative with the dupplet as $(\gamma,\beta)\in (\gamma<<0,\beta>>0)\cup (\gamma>>0,\beta<<0)$.

\subsection{Polytropic Fluid}
For the reconstruction of polytropic fluids using our present model, we have to bring the constraints as $A\rightarrow 0$ and $C\rightarrow 0$. Hence, the equation of state should become $p_f=(A\beta-\gamma)\rho_f^2$. The solutions can be written as follows.
\begin{equation}
    \rho_f=\frac{1}{C_1 a^3-(A\beta-\gamma)}
\end{equation}
\begin{equation}
    p_f=(A\beta-\gamma)(\frac{1}{C_1 a^3-(A\beta-\gamma)})^2
\end{equation}
and,
\begin{equation}
    \omega_f=\frac{(A\beta-\gamma)(\frac{1}{C_1 a^3-(A\beta-\gamma)})^2}{\frac{1}{C_1 a^3-(A\beta-\gamma)}}
\end{equation}
These solutions can provide an universe model which follows quadratic fluid formalism.

\section{Different Cosmological Phases and their fluid dynamics}
We aim to discuss different cosmological phases with parametric analysis from the new non linear model of our work. The model and its solutions are dependent upon four equation of state parameters and one power parameter. They are $A$, $\beta$, $\gamma$, $C$ and $\alpha$ respectively. Now the magnitudes of those parameters can provide different cosmological phases, their evolutions and smooth phase transitions. Here, we have discussed the parametric analysis for getting some cosmic phases and their transformations in table 1.
\begin{table}[ht]
    \centering
    \begin{tabular}{|c|c|}
    \hline
       \textbf{Cosmic Phases}  &  \textbf{Parametric Analysis}\\
       \hline
       \hline
      \small{Phantom Phase}  & \small{ $A<0$, $B<0$ and $C>0$;} \\
         & \small{ hence, $\beta>0$ and $\gamma>0$ with $\alpha<0$}\\
       \hline
       \small{Quintom Transformation } & \small{ $A>0$, $B>0$ and $C>>0$;}\\
         & \small{ hence, either $\beta>>0$ or $\gamma<<0$ with $\alpha>0$ or $\alpha<1$}\\
       \hline
      \small{Quintessence Phase } & \small{ $A>0$, $B>0$ and $C>>0$;}\\
         & \small{ hence, both $\beta>>0$ and $\gamma<<0$ with $\alpha>>0$}\\
       \hline
      \small{CDM Dominant Phase } & \small{$A=0$, $B=0$ and $C=0$; }\\
         & \small{ hence, $A\beta=\gamma$ with $\alpha=$ arbitrary} \\
       \hline
      \small{WDM Dominant Phase with Radiation Region } & \small{ $A>0$, $B>0$ and $C<<0$ as well as $C<0$;} \\
         & \small{ as well as $C<0$;} \\
         &  \small{hence, both $\beta>>0$ and $\gamma<<0$; or also }\\
         & \small{ hence, either $\beta>>0$ or $\gamma<<0$ with $\alpha>0$}\\
       \hline 
    \end{tabular}
    \caption{Parametric Analysis of different Cosmological Phases}
    \label{tab:my_label1}
\end{table}
The table 1 represents the region of the magnitudes of the parameters which can be used to discuss different cosmic phases and their transformations as well as their evolutions. We have shown the densities and pressures for those phases graphically in next sections. We have chosen the values of the parameters to plot the figures according to the region derived in table 1. $\alpha$ has been considered positive to bring the end of each phase with evolution w.r.t. scale factor.

\section{Result Analysis and Comparative studies}
We have used pre predicted magnitudes of the parameters and our results is capable of establishing the Phantom phase, Quintom region, Quintessence phase, Cold Dark Matter or CDM phase as well as Warm Dark matter i.e. WDM and radiation properly with exact values of the equation of state parameters. The results has been given in figure 1 from a to o. We have given the results for different phases in different graphs. All the densities are positives in magnitudes and pressures are negative except for the radiation phase. For Phantom we have found $\omega_f<-1$, for quintom its around $\omega_f\approx-1$. For Quintessence and CDM the equation of state parameter gave $0>\omega_f>-1$ and $\omega_f=0$ respectively. For radiation, both pressure and density provides positive values and hence, the equation of state parameter comes as $\omega_f\approx0.33$. All these results found are in their evolving state i.e. all the energy densities and pressures are changing with scale factor. Hence, we could provide smooth transitions during the phase transitions of universe. In our analysis, we consider only the fluid dynamics and our results are in agreement with the experimental results. Except the Phantom dominant universe, we found $\left|\rho_f\right|>\left|p_f\right|$. Hence, these models are stable with Null energy conditions, Weak energy conditions as well as Dominant energy conditions in the perspective of fluid dynamics. All of these models provide $\left|\rho_f\right|<<\left|3p_f\right|$ and thus, with the violation of strong energy conditions they can provide accelerated expanding universe solutions.\par
The results shown in figure 1 provide specific region of the scale factors where they can only provide the cosmological phases discussed here. These phenomena strictly proved that no cosmic phases discussed here, can be stable for permanent period of time. All of them must start and end through some phase transitions. These phase transitions should be controlled by the four cosmic parameters and one power parameters discussed in our model. Thus, we can conclude that the parameters of our model are also time varying during the phase transitions but constant or almost constant during specific phases. The change of the magnitudes of those parameters provide an idea about the order of phase transitions for those phases.

\begin{figure}[H]
\begin{subfigure}{.33\textwidth}
  \centering
  \includegraphics[width=0.8\linewidth]{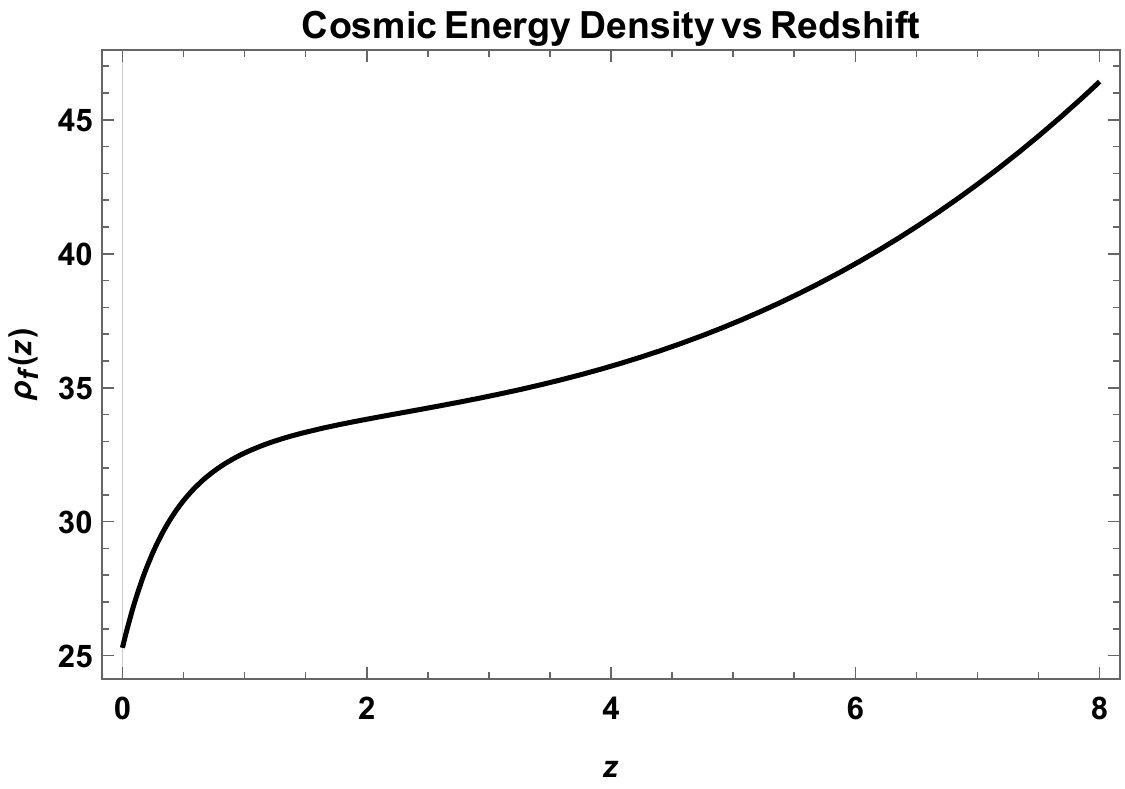}  
  \caption{$\rho_{f}$ vs $z$}
  \label{fig:sub-first}
\end{subfigure}
\begin{subfigure}{.33\textwidth}
  \centering
  \includegraphics[width=.8\linewidth]{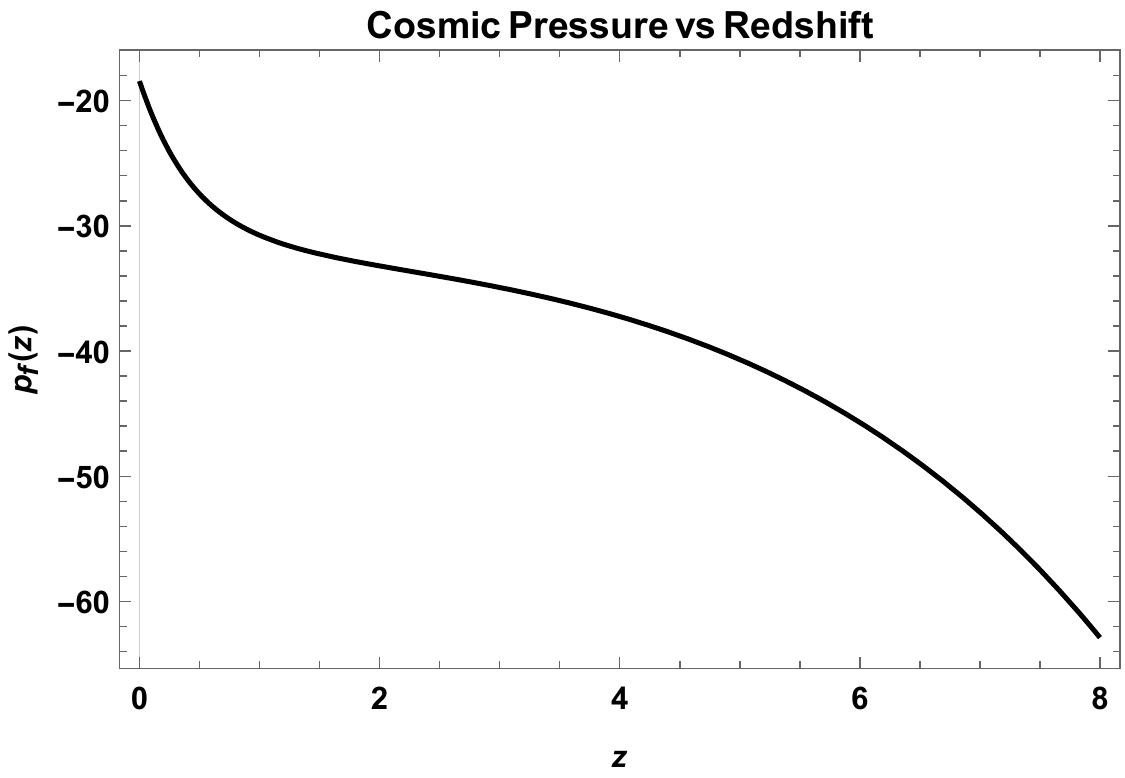}  
  \caption{$p_{f}$ vs $z$}
  \label{fig:sub-second}
\end{subfigure}
\begin{subfigure}{.33\textwidth}
  \centering
  \includegraphics[width=.8\linewidth]{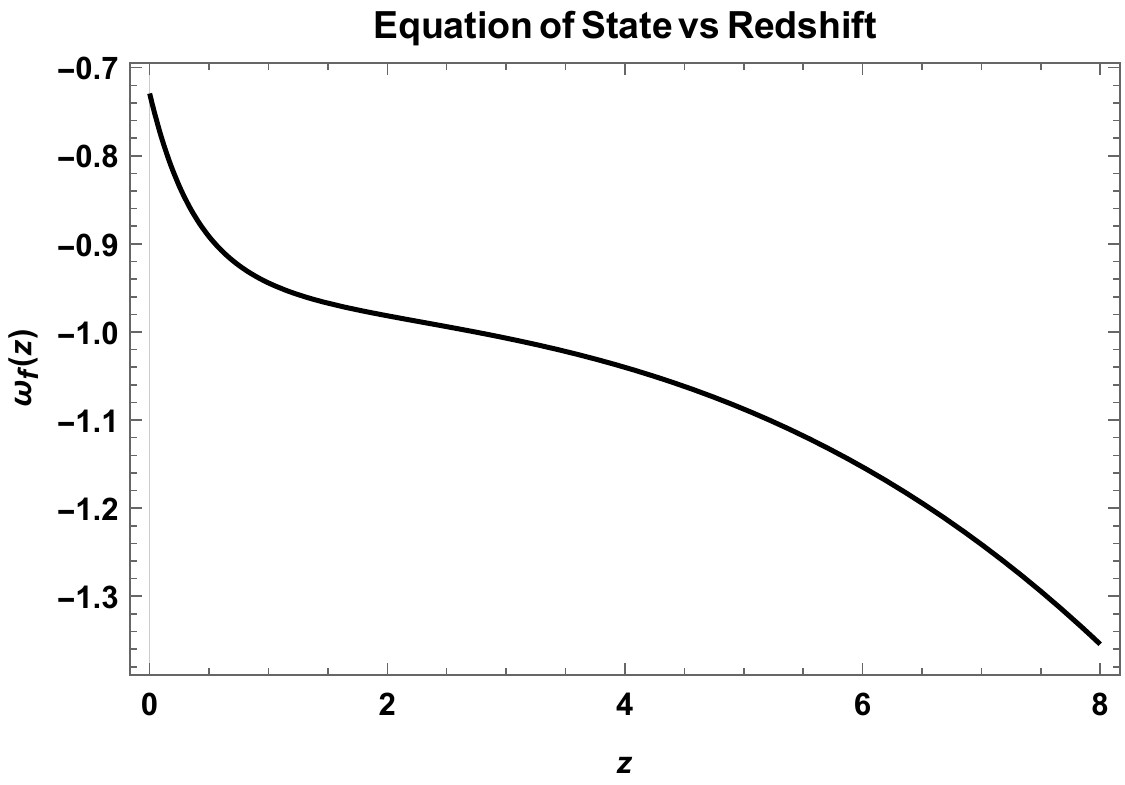}  
  \caption{$\omega_f=\frac{p_f}{\rho_f}$ vs $z$}
  \label{fig:sub-third}
\end{subfigure}


\begin{subfigure}{.33\textwidth}
  \centering
  \includegraphics[width=.8\linewidth]{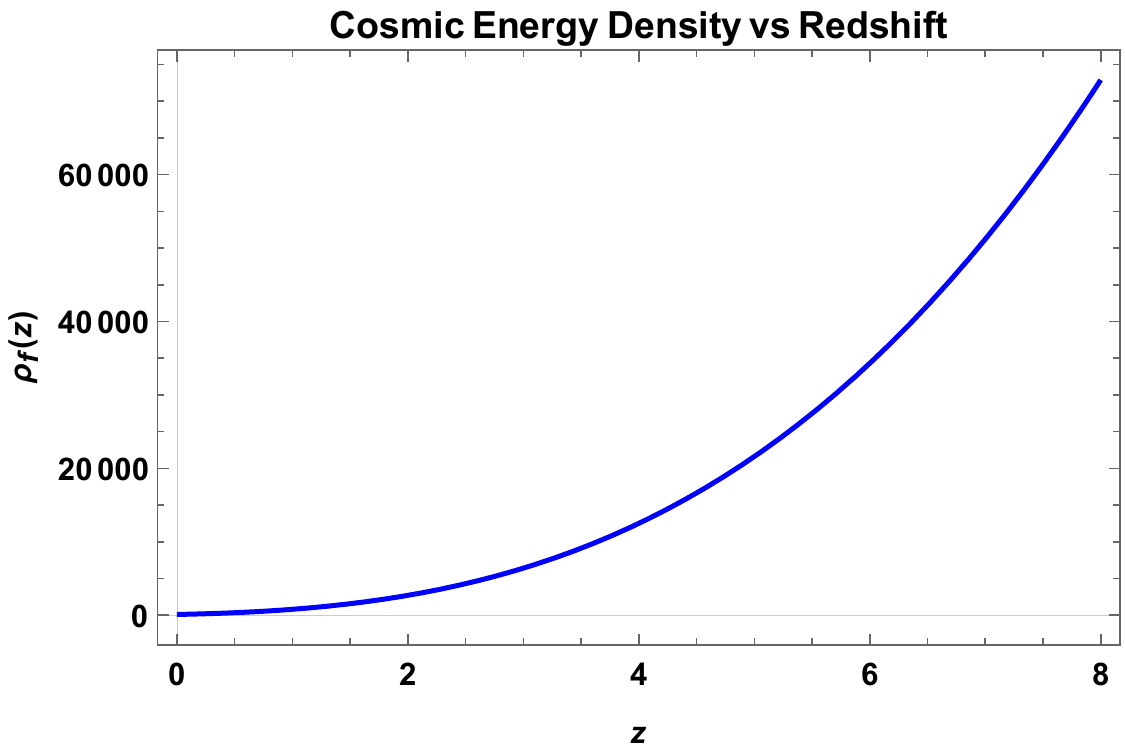}  
  \caption{$\rho_{f}$ vs $z$}
  \label{fig:sub-fourth}
\end{subfigure}
\begin{subfigure}{.33\textwidth}
  \centering
  \includegraphics[width=.8\linewidth]{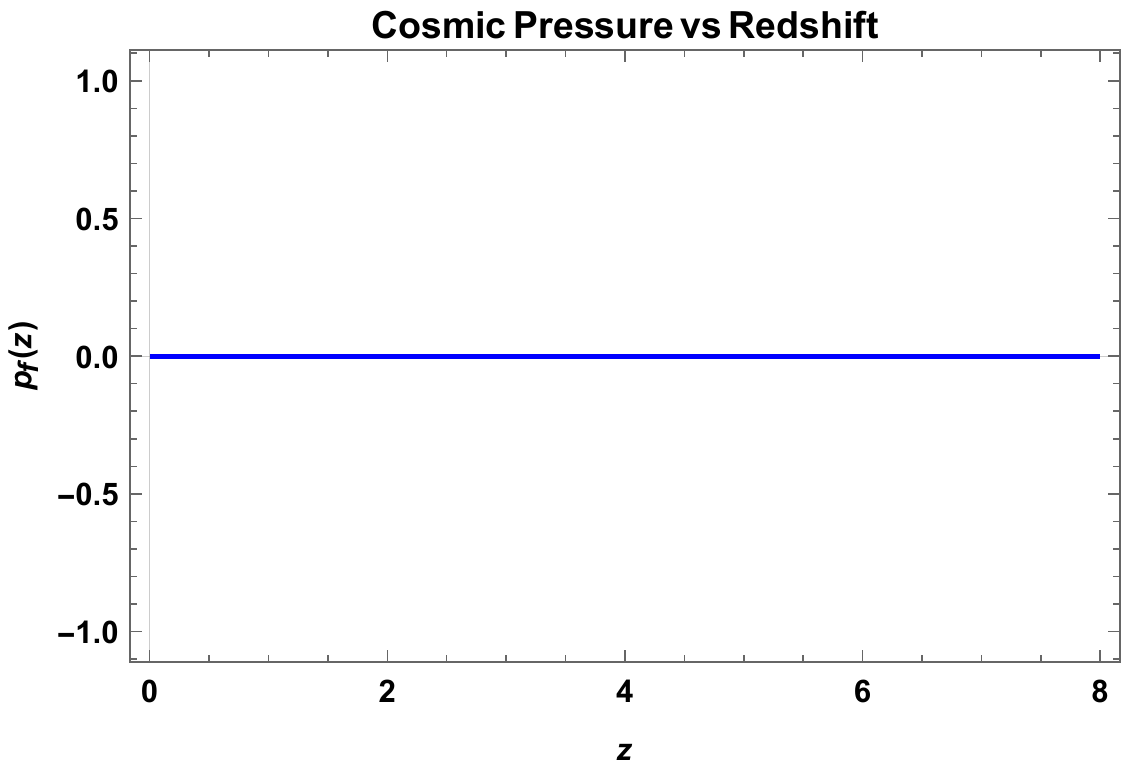}  
  \caption{$p_{f}$ vs $z$}
  \label{fig:sub-fifth}
\end{subfigure}
\begin{subfigure}{.33\textwidth}
  \centering
  \includegraphics[width=.8\linewidth]{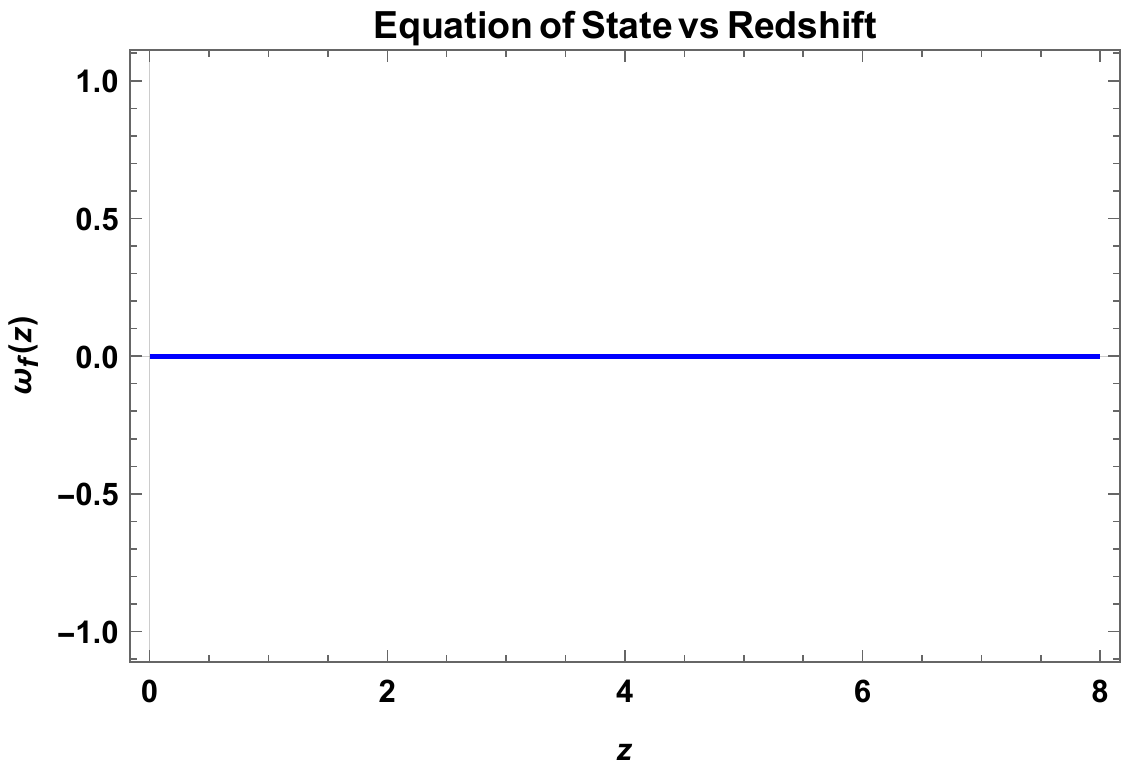}  
  \caption{$\omega_f=\frac{p_f}{\rho_f}$ vs $z$}
  \label{fig:sub-sixth}
\end{subfigure}


\begin{subfigure}{.33\textwidth}
  \centering
  \includegraphics[width=0.8\linewidth]{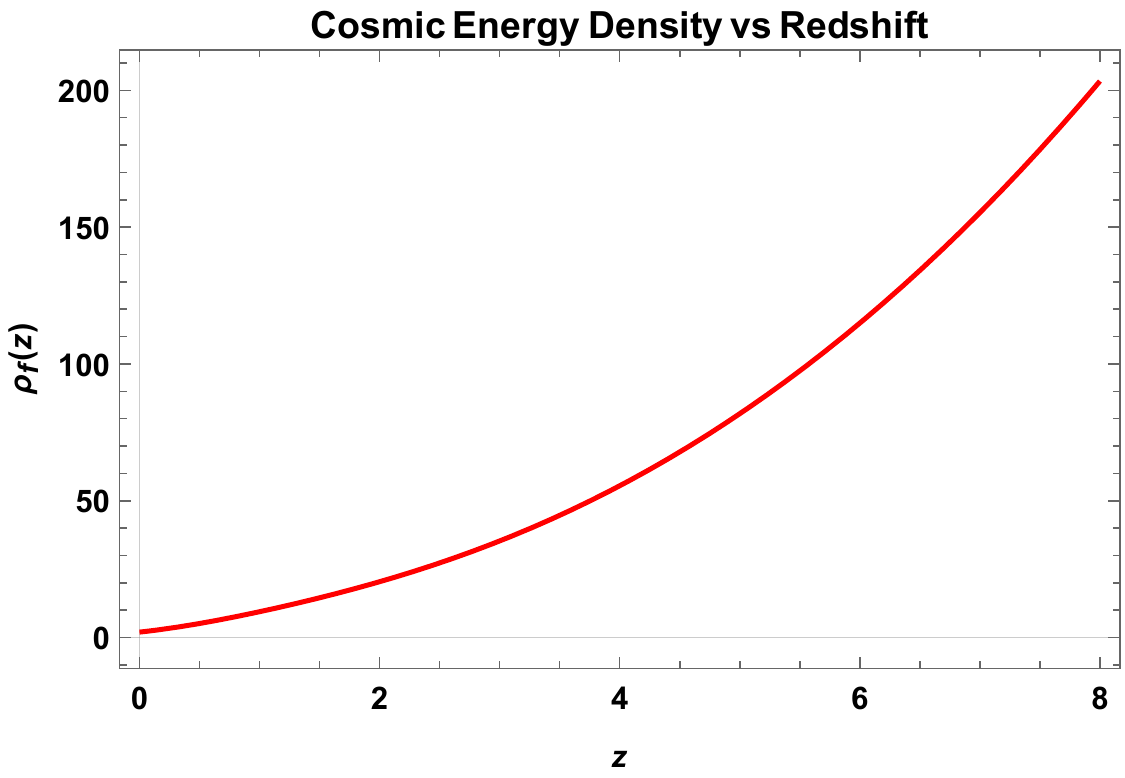}  
  \caption{$\rho_{f}$ vs $z$}
  \label{fig:sub-seventh}
\end{subfigure}
\begin{subfigure}{.33\textwidth}
  \centering
  \includegraphics[width=.8\linewidth]{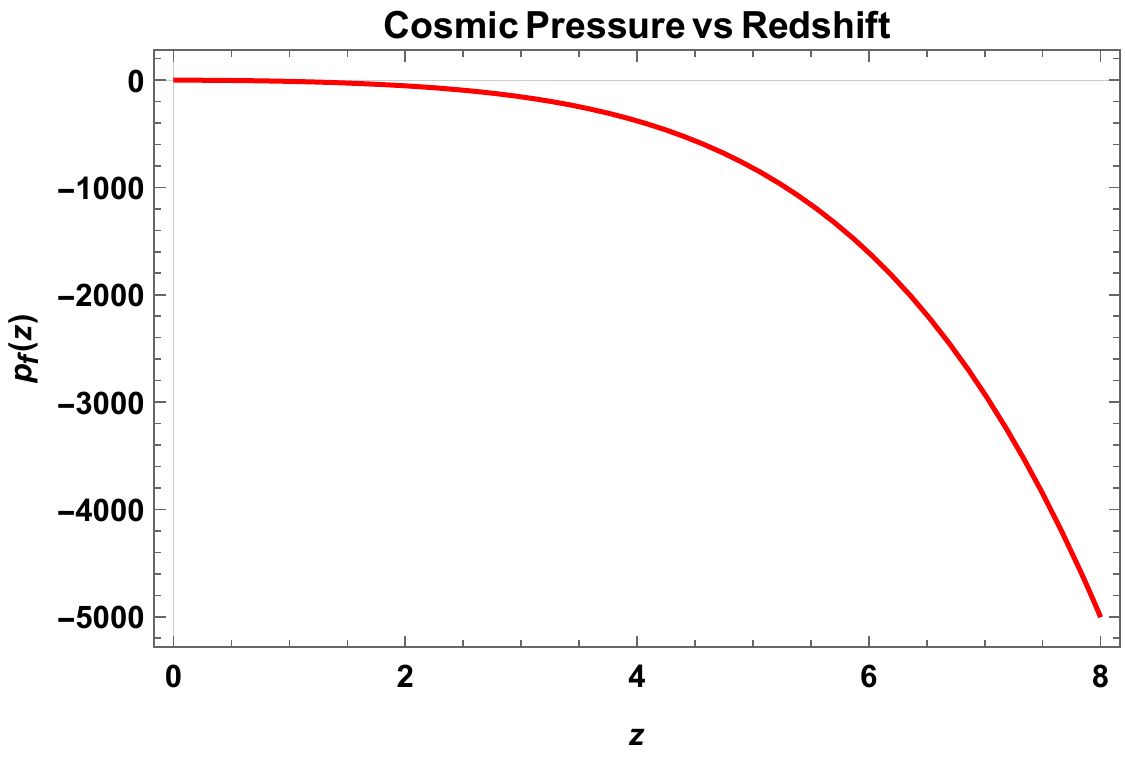}  
  \caption{$p_{f}$ vs $z$}
  \label{fig:sub-eighth}
\end{subfigure}
\begin{subfigure}{.33\textwidth}
  \centering
  \includegraphics[width=.8\linewidth]{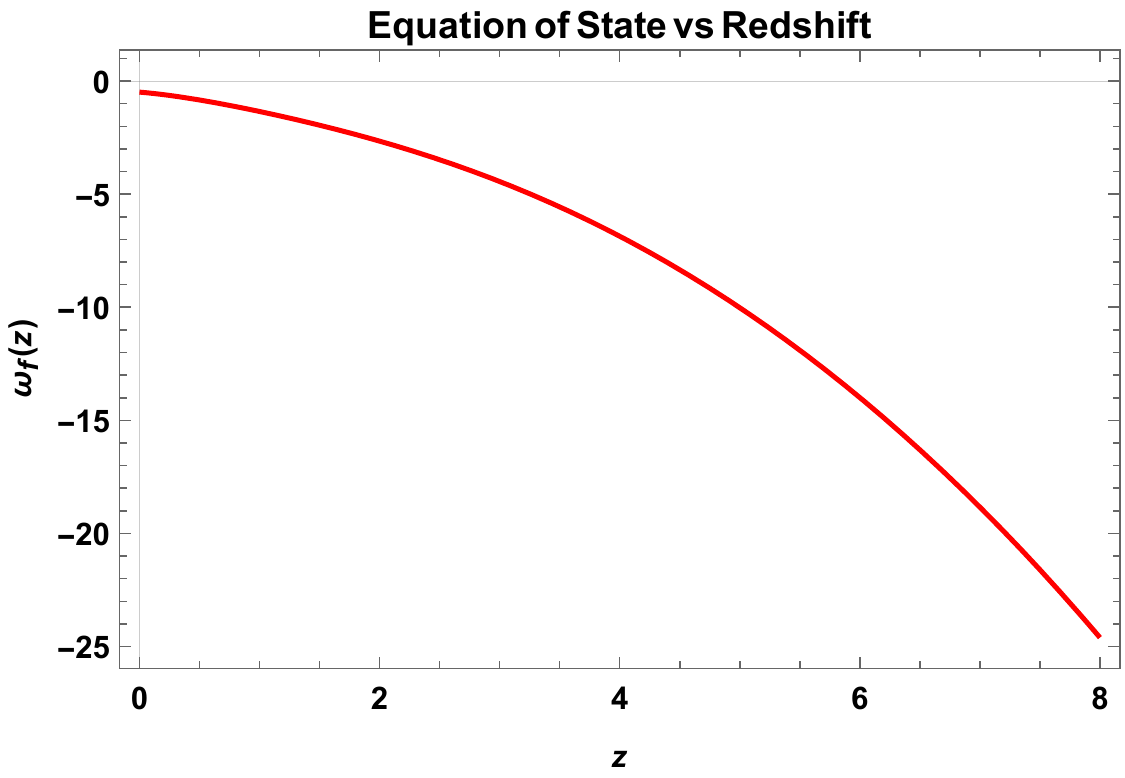}  
  \caption{$\omega_f=\frac{p_f}{\rho_f}$ vs $z$}
  \label{fig:sub-ninth}
\end{subfigure}


\begin{subfigure}{.33\textwidth}
  \centering
  \includegraphics[width=.8\linewidth]{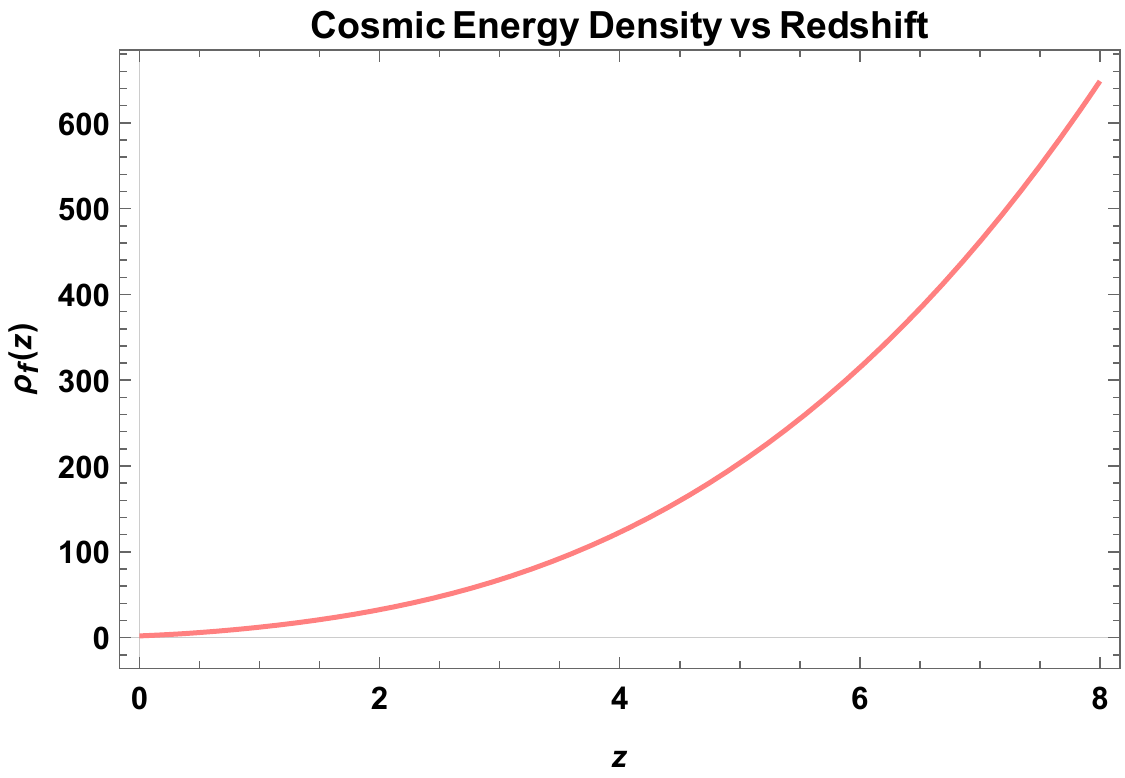}  
  \caption{$\rho_{f}$ vs $z$}
  \label{fig:sub-tenth}
\end{subfigure}
\begin{subfigure}{.33\textwidth}
  \centering
  \includegraphics[width=.8\linewidth]{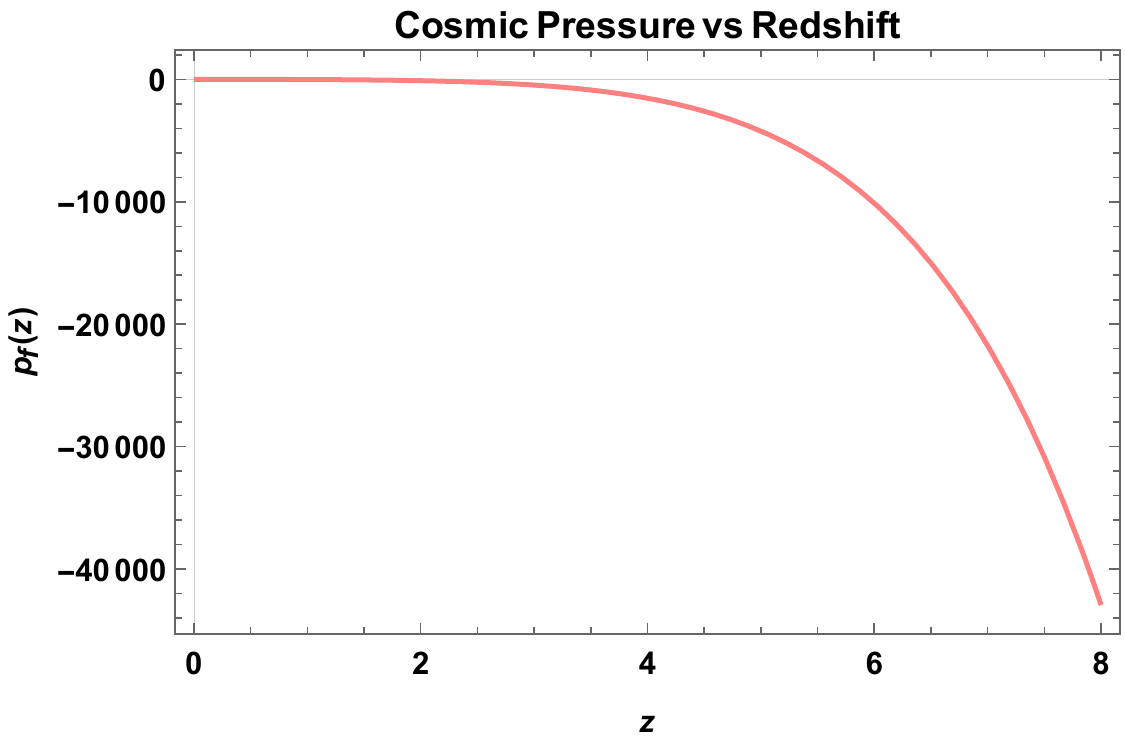}  
  \caption{$p_{f}$ vs $z$}
  \label{fig:sub-eleventh}
\end{subfigure}
\begin{subfigure}{.33\textwidth}
  \centering
  \includegraphics[width=.8\linewidth]{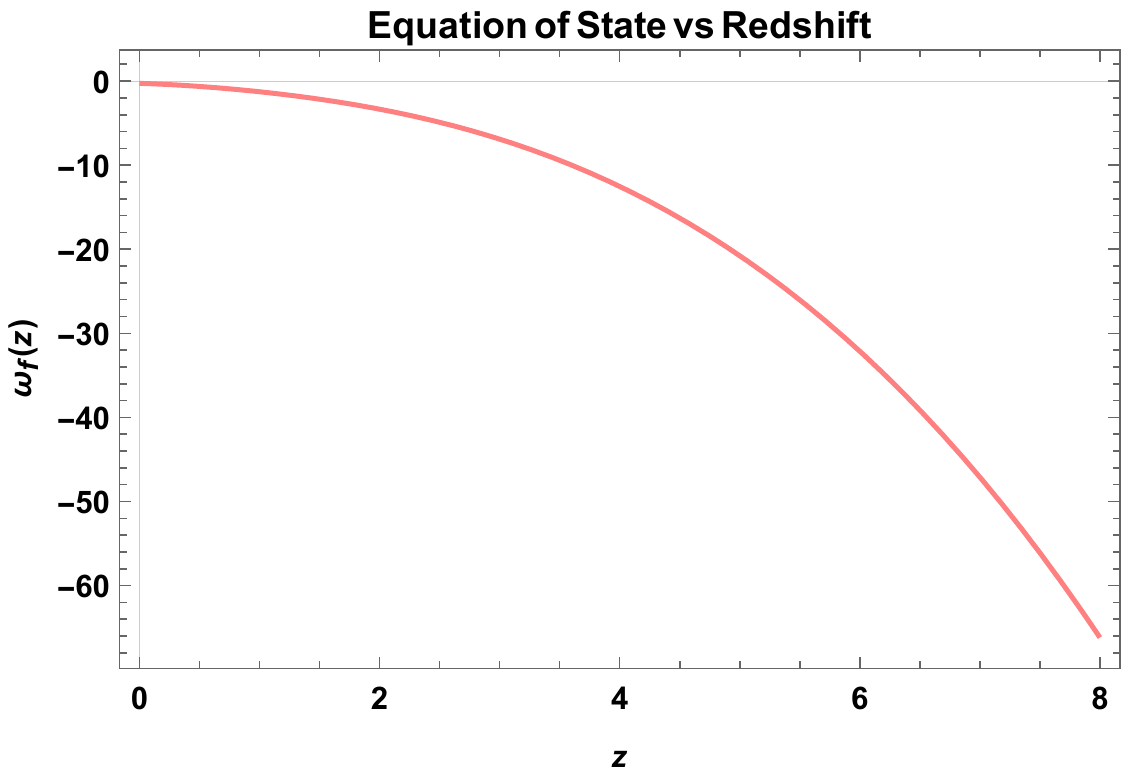}  
  \caption{$\omega_f=\frac{p_f}{\rho_f}$ vs $z$}
  \label{fig:sub-twelfth}
\end{subfigure}


\begin{subfigure}{.33\textwidth}
  \centering
  \includegraphics[width=.8\linewidth]{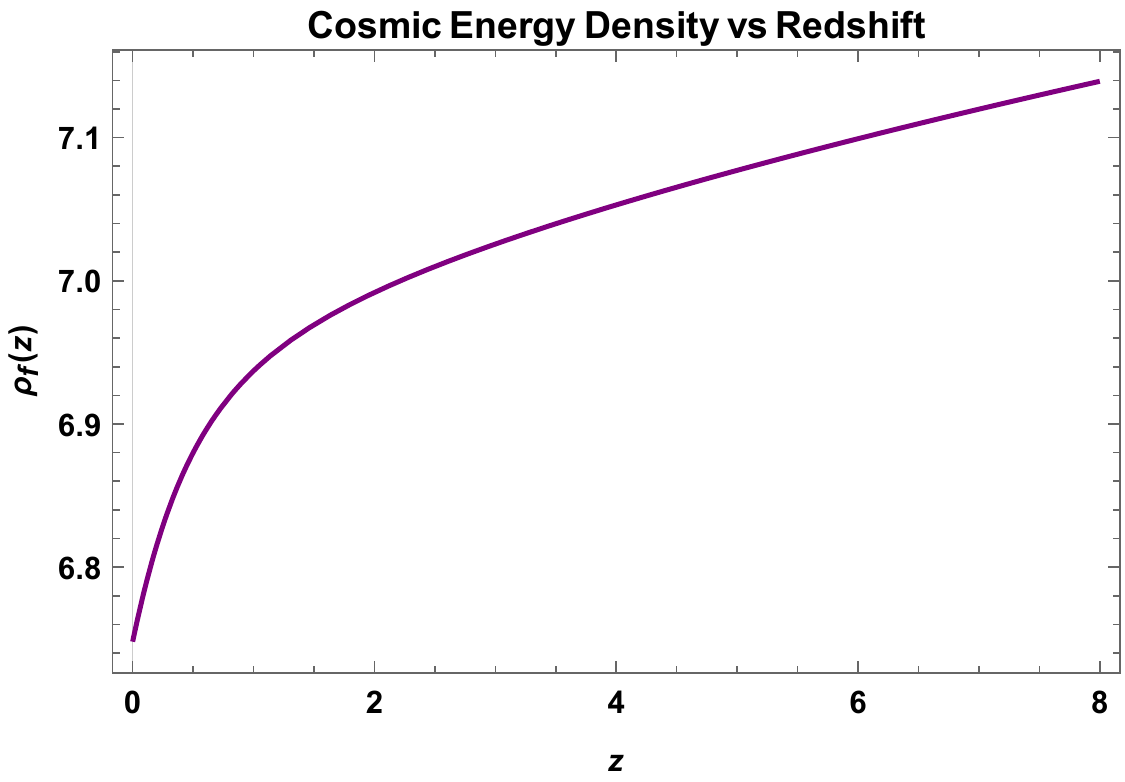}  
  \caption{$\rho_{f}$ vs $z$}
  \label{fig:sub-thirteenth}
\end{subfigure}
\begin{subfigure}{.33\textwidth}
  \centering
  \includegraphics[width=.8\linewidth]{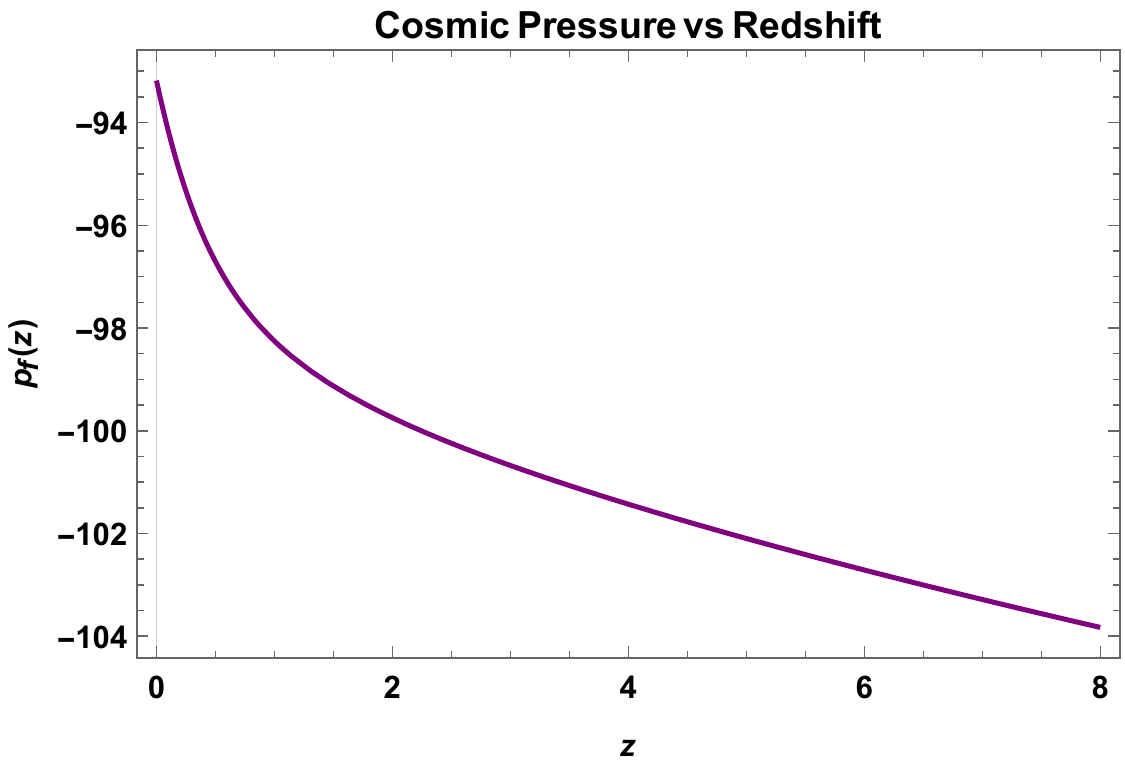}  
  \caption{$p_{f}$ vs $z$}
  \label{fig:sub-fourteenth}
\end{subfigure}
\begin{subfigure}{.33\textwidth}
  \centering
  \includegraphics[width=.8\linewidth]{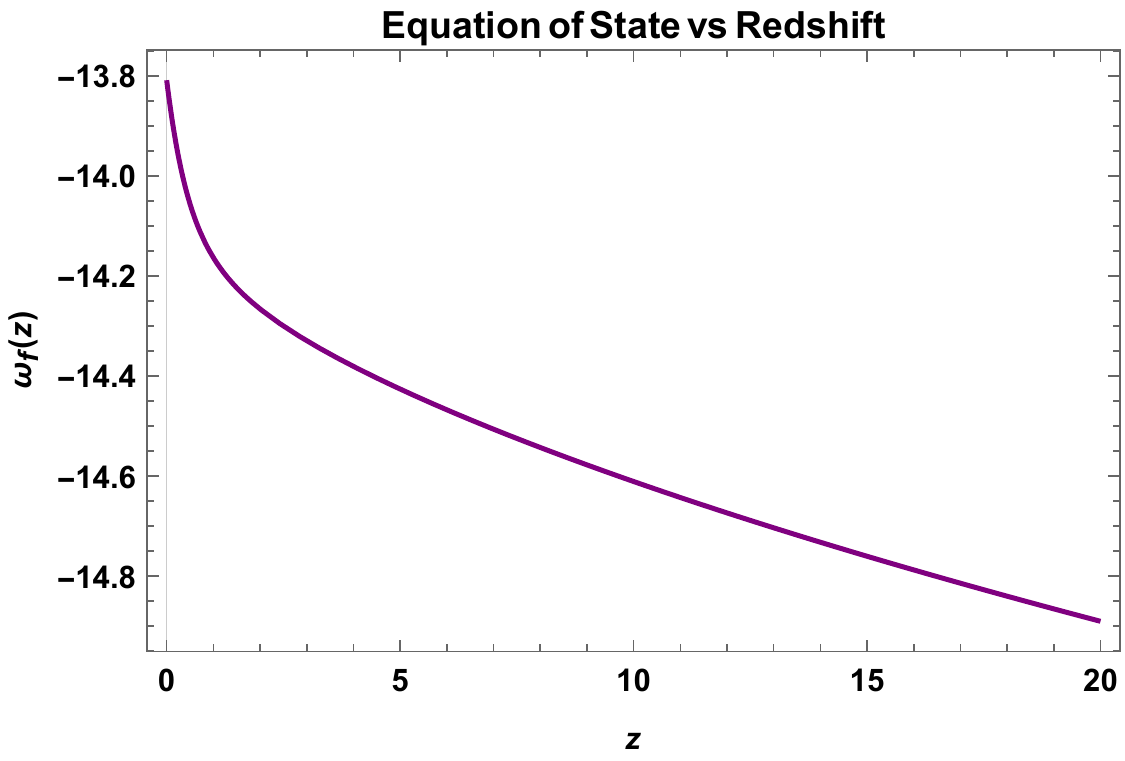}  
  \caption{$\omega_f=\frac{p_f}{\rho_f}$ vs $z$}
  \label{fig:sub-fifteenth}
\end{subfigure}

\caption{Black for Radiation and WDM, Blue for CDM, Red for Quintessence, Pink for Quintom and Purple for Phantom universe}
\label{fig:fig}
\end{figure}

\section{Conclusion}
We have finally, established the new non linear cosmological fluid model through the detailed discussion of fluid dynamics. The whole discussion proceeds with the scale factor variation of energy density and pressure for a new non linear equation of state under Einsteinian gravity action with non varying gravitational constant. We have matched our results with the well established values of the equation of state parameter for different cosmological phases. Our results also violate strong energy condition to provide acceleration in the expansion scheme of our model. The null energy condition for all those phases is satisfied, because we found $\left|\rho_f\right|>\left|p_f\right|$. Hence, phenomenologically, our model obeys the first law of thermodynamics beyond phantomic region of universe. The details of the thermodynamical analysis is beyond the scope of this work.

\begin{figure}[H]
\begin{subfigure}{.5\textwidth}
  \centering
  \includegraphics[width=0.8\linewidth]{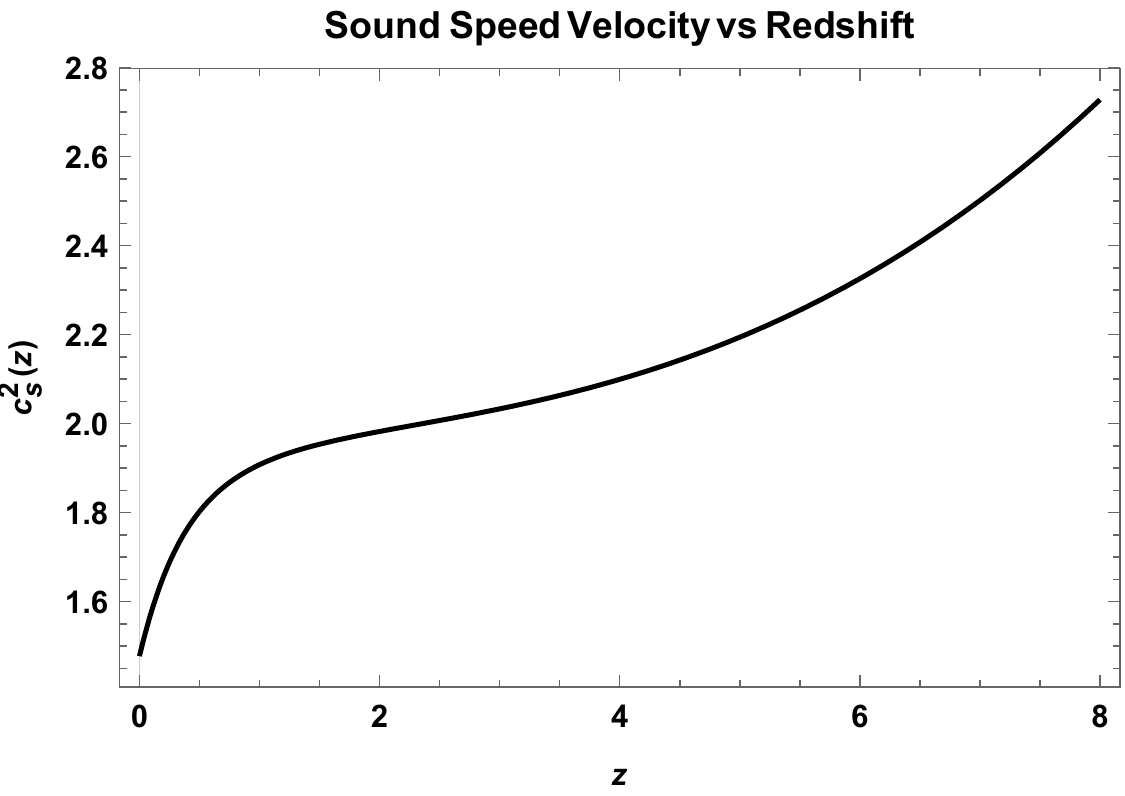}  
  \caption{$c_{s}^2$ vs $z$}
  \label{fig:sub-16}
\end{subfigure}
\begin{subfigure}{.5\textwidth}
  \centering
  \includegraphics[width=.8\linewidth]{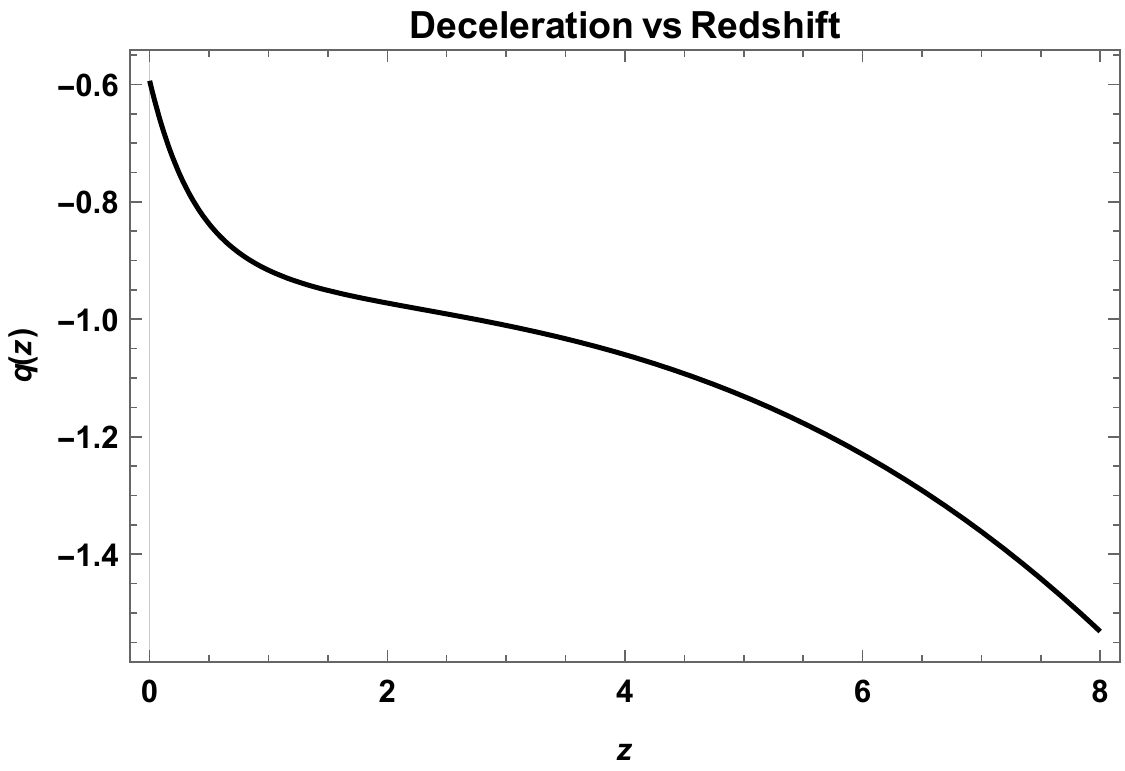}  
  \caption{$q(z)$ vs $z$}
  \label{fig:sub-17}
\end{subfigure}


\begin{subfigure}{.5\textwidth}
  \centering
  \includegraphics[width=.8\linewidth]{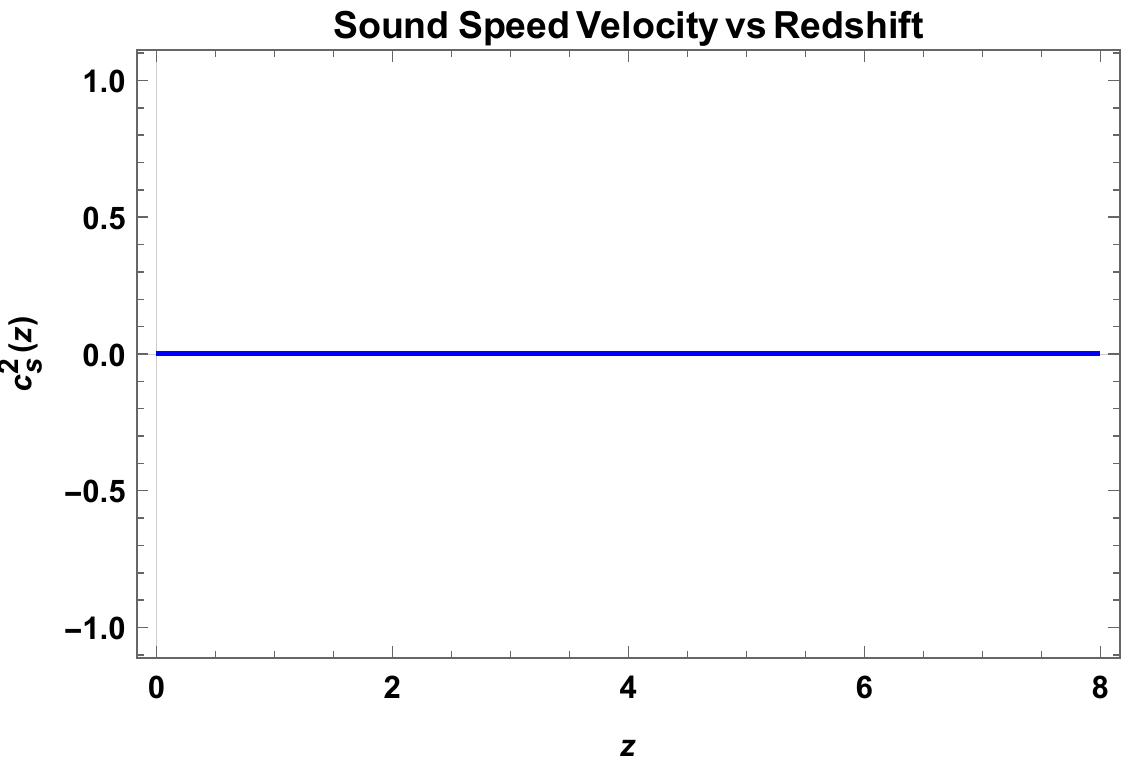}  
  \caption{$c_{s}^2$ vs $z$}
  \label{fig:sub-18}
\end{subfigure}
\begin{subfigure}{.5\textwidth}
  \centering
  \includegraphics[width=.8\linewidth]{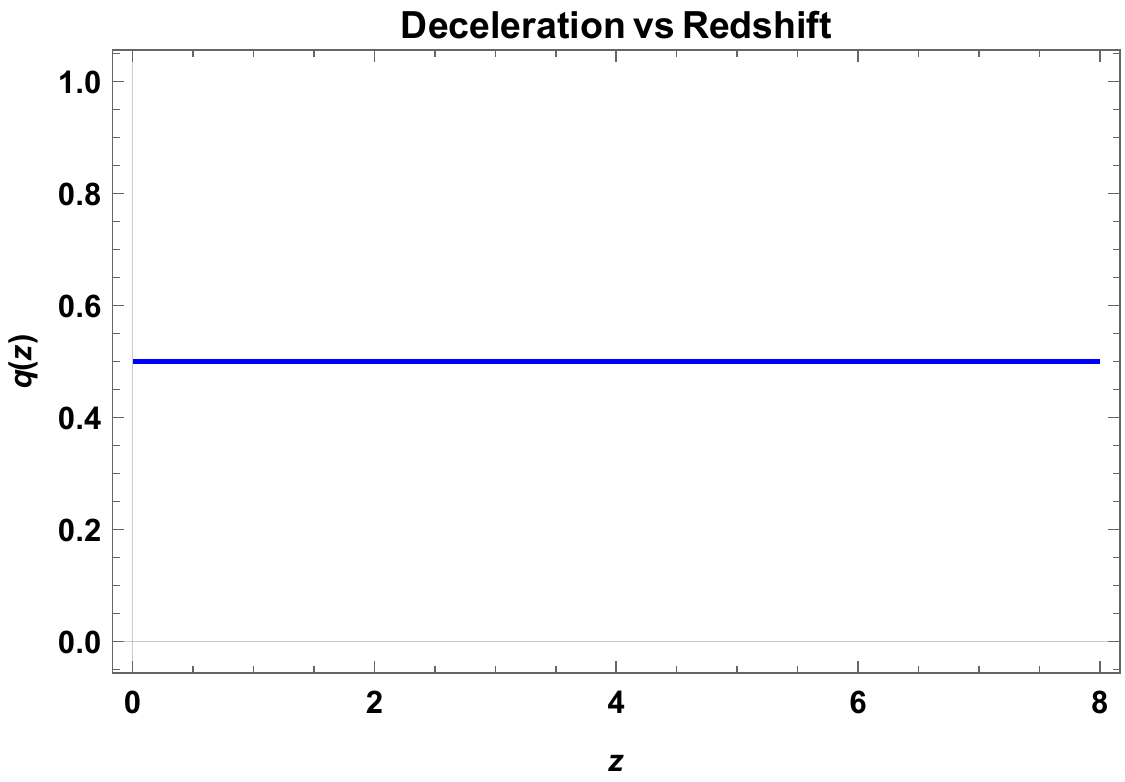}  
  \caption{$q(z)$ vs $z$}
  \label{fig:sub-19}
\end{subfigure}


\begin{subfigure}{.5\textwidth}
  \centering
  \includegraphics[width=0.8\linewidth]{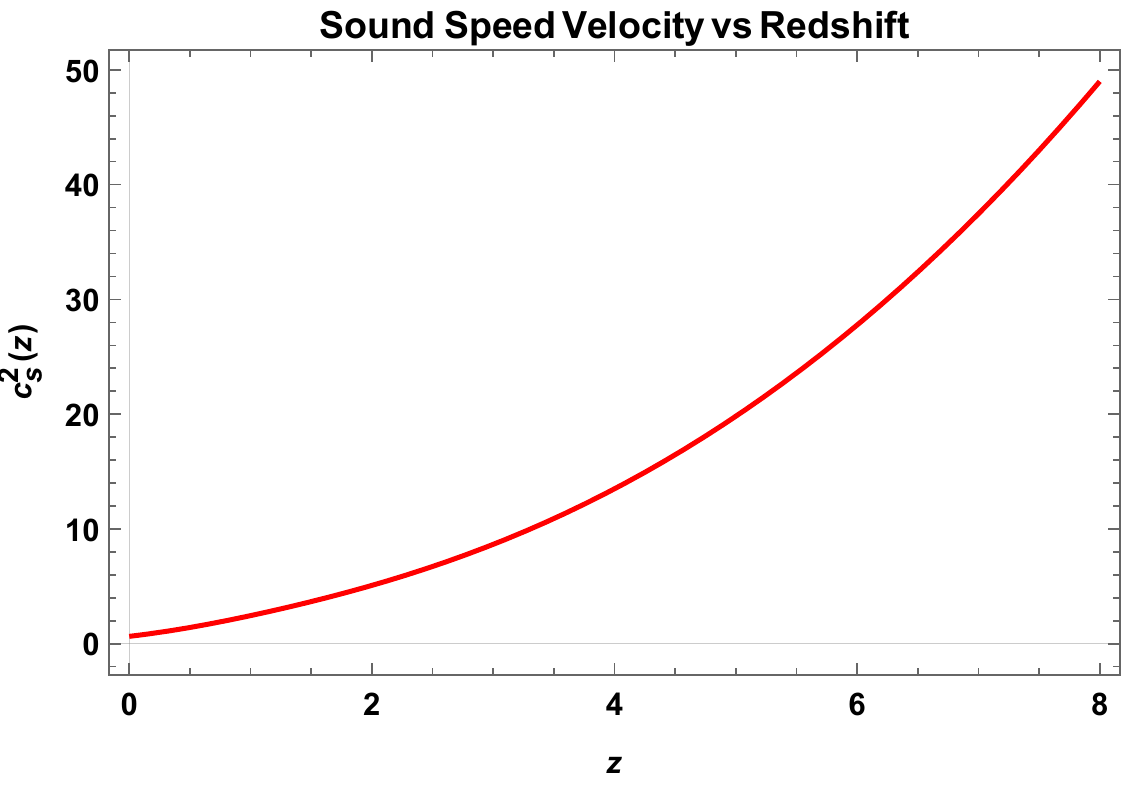}  
  \caption{$c_{s}^2$ vs $z$}
  \label{fig:sub-20}
\end{subfigure}
\begin{subfigure}{.5\textwidth}
  \centering
  \includegraphics[width=.8\linewidth]{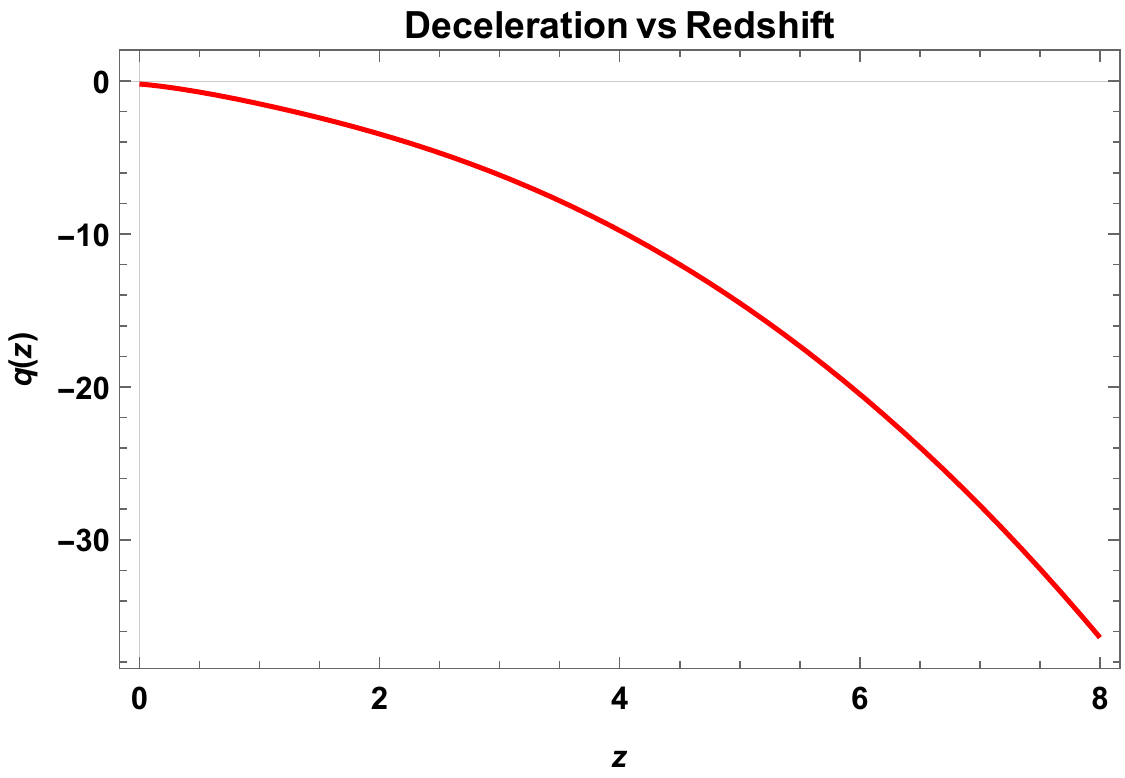}  
  \caption{$q(z)$ vs $z$}
  \label{fig:sub-21}
\end{subfigure}

\caption{Black for Radiation and WDM, Blue for CDM, Red for Quintessence universe}
\label{fig:fig}
\end{figure}

\begin{figure}[H]
\begin{subfigure}{.5\textwidth}
  \centering
  \includegraphics[width=.8\linewidth]{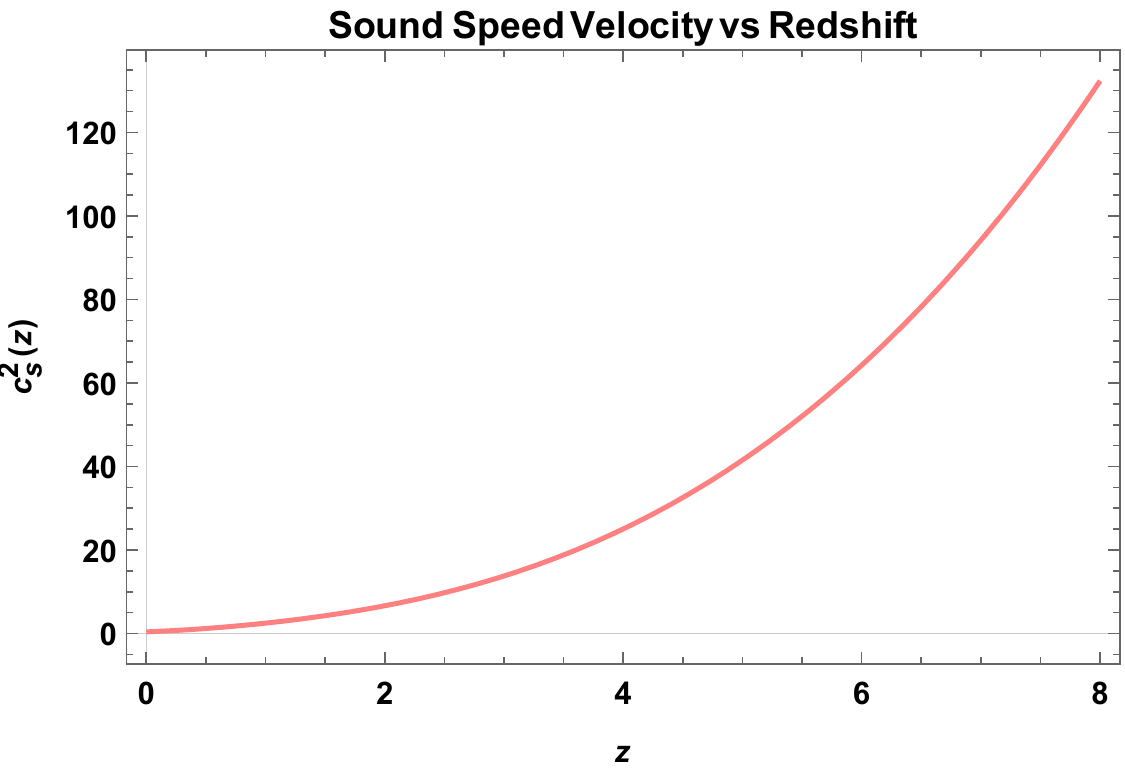}  
  \caption{$c_{s}^2$ vs $z$}
  \label{fig:sub-22}
\end{subfigure}
\begin{subfigure}{.5\textwidth}
  \centering
  \includegraphics[width=.8\linewidth]{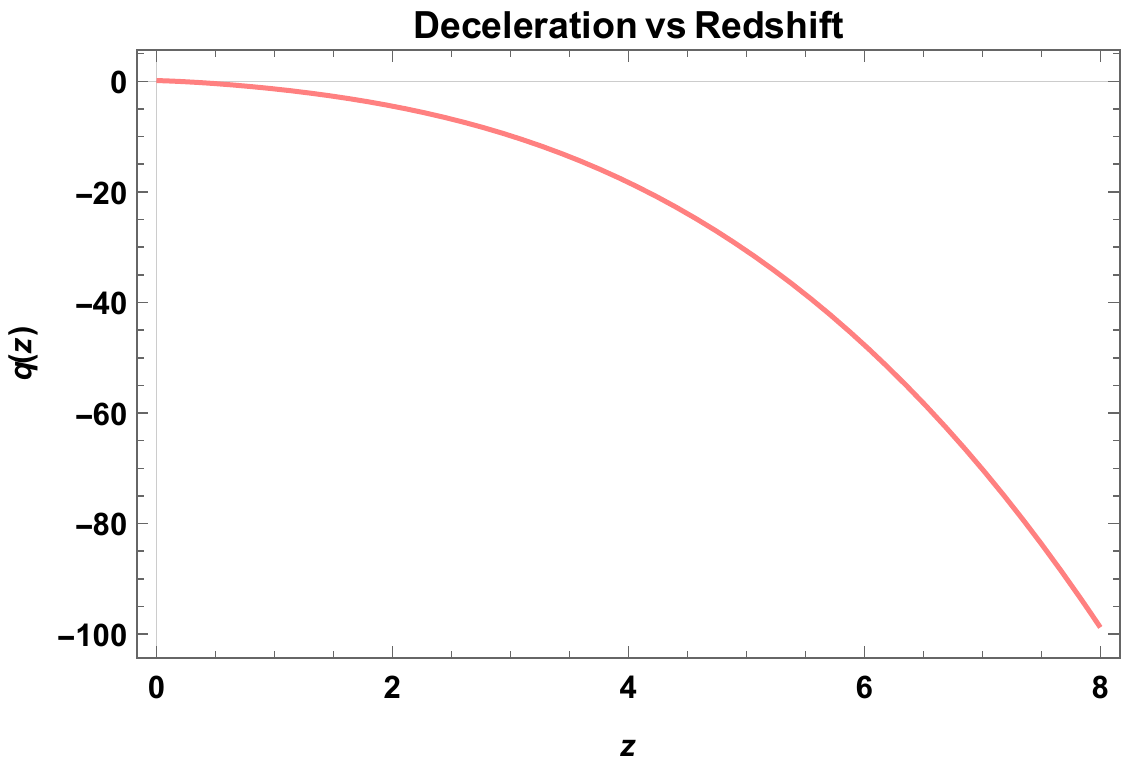}  
  \caption{$q(z)$ vs $z$}
  \label{fig:sub-23}
\end{subfigure}


\begin{subfigure}{.5\textwidth}
  \centering
  \includegraphics[width=.8\linewidth]{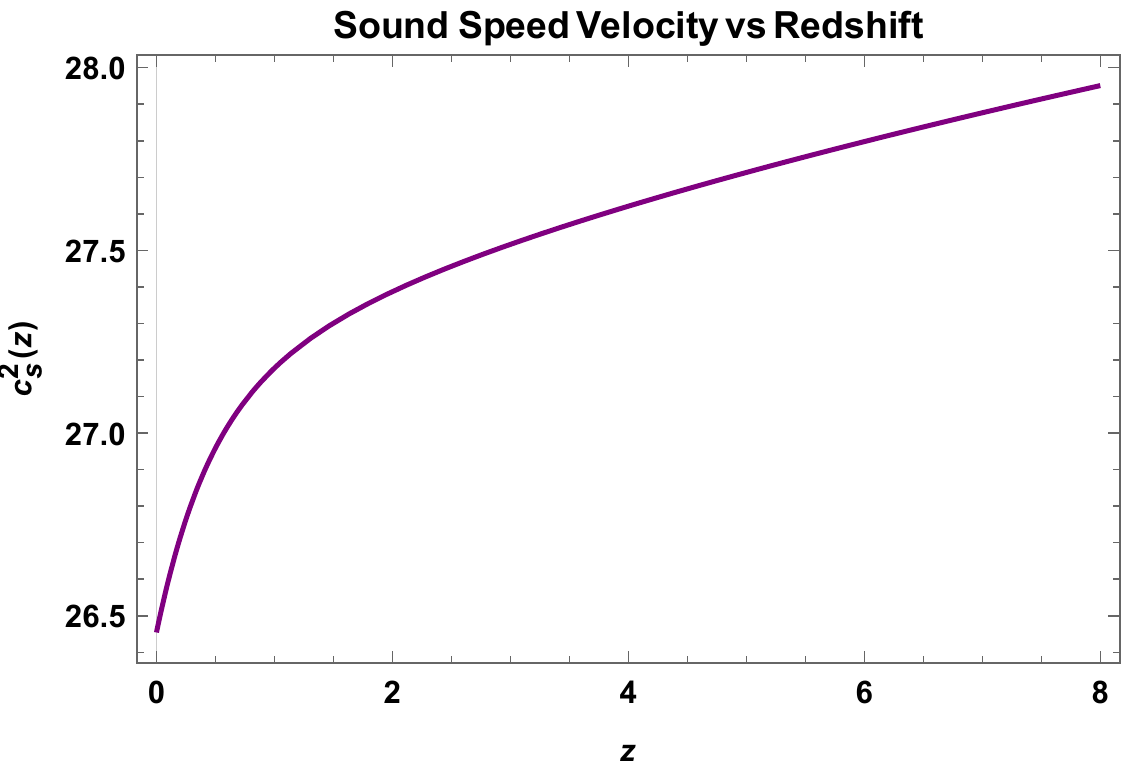}  
  \caption{$c_{s}^2$ vs $z$}
  \label{fig:sub-24}
\end{subfigure}
\begin{subfigure}{.5\textwidth}
  \centering
  \includegraphics[width=.8\linewidth]{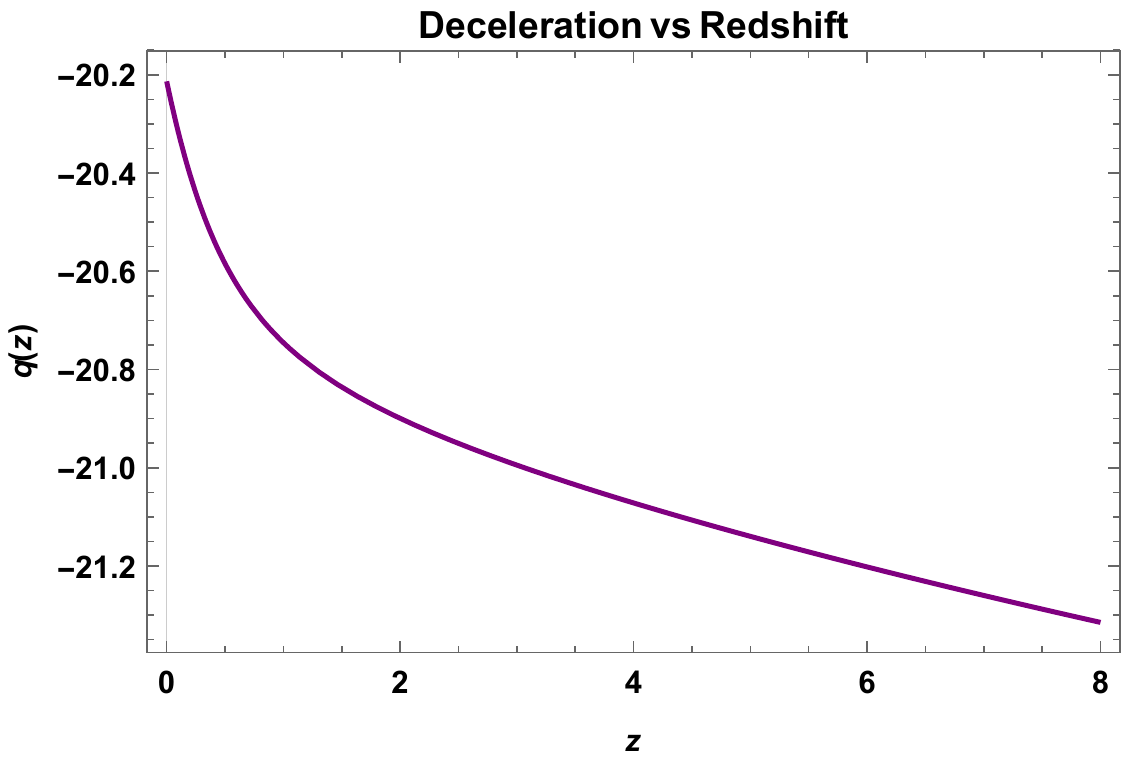}  
  \caption{$q(z)$ vs $z$}
  \label{fig:sub-25}
\end{subfigure}

\caption{Pink for Quintom and Purple for Phantom universe}
\label{fig:fig}
\end{figure}

The variation of deceleration parameters and sound wave velocities have been presented in 10 graphs of figures 2 and 3. Those graphs have been given to represent the stability of our model discussed in this paper. The sound wave velocity should be positive when the model is stable under scalar perturbation. Hence this four parameter equation of state based cosmic fluid model is stable under perturbation. The deceleration parameters for each phase provide negative values which proved that all those models are representing cosmic acceleration.

\section*{Acknowledgement}
Surajit Chattopadhyay acknowledges financial support from CSIR under the Grant No.03(1420) / 18 / EMR-II. The authors are thankful to the editors and reviewers for their insightful comments for valuable improvements to our paper.

\section*{Appendix}
The generalized Van--Der-Waals model or VDW fluid equation of state for cosmic dynamics can be written as follows.
\begin{equation}
    p_f=\frac{A\rho_f}{1-\beta\rho_f}-\gamma\rho_f^2
\end{equation}
Now this equation of state can be expanded into binomial series as follows. The denominator of the first term of RHS of the above equation has been expanded with binomial series.
\begin{equation}
    p_f=A\rho_f(1+\beta \rho_f+(\beta\rho_f)^2+(\beta \rho_f)^3+....)-\gamma\rho_f^2
\end{equation}
Now, if we consider $\left |\beta \rho_f\right |<1$, we can converge the expanded equation of state as 
\begin{equation}
    p_f=A\rho_f+(A\beta-\gamma)\rho_f^2
\end{equation}
Now in presence of that condition, we can provide the following analysis which can control the cosmic dynamics under that equation of state. The condition $\left |\beta\rho_f\right |<1$ can be extended into following inequality.
\begin{equation}
    \left |\beta \rho_f\right | < 1 \rightarrow -1 < \beta \rho_f < 1
\end{equation}
Hence, we can provide two concluding statements which are as follows. The mathematical form of the first statement is as follows.
\begin{equation}
    -\frac{1}{\beta} < \rho_f < \frac{1}{\beta}
\end{equation}
and, the second statement is as follows.
\begin{equation}
    -\frac{1}{\rho_f} < \beta < \frac{1}{\rho_f}
\end{equation}
The first statement (equation 40) represents the boundary condition of energy density. For having fixed parameter $\beta$ there must be some upper and lower boundary for energy density in the cosmic fluid system under the specific type of non-linear model or specific VDW model. Beyond this limits the cosmic fluid will get some phase transition and hence it starts to follow different equation of states. \\
The second statement provides the nature of variation of parameter $\beta$. According to the representation the parameter $\beta$ should vary hyperbolically with energy density with a limit of $(-\frac{1}{\rho_f},\frac{1}{\rho_f})$. Whenever the parameter crosses the limit, the universe should get some phase transition and the cosmic fluid starts to follow different equation of states. Another possibility is that the cosmic phase should get one bounce in terms of parameter. The values of the parameter should return to limit and the universe faces some new cosmological era. In this case, the change of equation of state is not a necessary phenomena for further expansion of universe.\\
In the present context we have followed the first statement because the first statement can give more clearer explanation of the cosmic phase transition dynamics.

\end{document}